\documentclass[12pt]{article}
\usepackage{graphicx}
\begin{document}

\begin{center}

\vbox{\vspace{10mm}}
{\LARGE \bf  Symmetries shared by the Poincar\'e Group and the Poincar\'e Sphere}

\vspace{10mm}

Young S. Kim\footnote{electronic mail: yskim@umd.edu} \\
Department of Physics, University of Maryland, College Park,
Maryland 20742 \\

\vspace{5mm}

Marilyn E. Noz \footnote{electronic mail: marilyne.noz@gmail.com} \\
Department of Radiology, New York University, New York, New York
10016

\end{center}

\vspace{10mm}

\abstract{Henri Poincar\'e formulated the mathematics of Lorentz
transformations, known as the Poincar\'e group.  He also formulated the
Poincar\'e sphere for polarization optics.  It is shown that these two
mathematical instruments can be derived from the two-by-two representations
of the Lorentz group.  Wigner's little groups for internal space-time
symmetries are studied in detail.  While the particle mass is a
Lorentz-invariant quantity, it is shown possible to address its variations
in terms of the decoherence mechanism in polarization optics.}

\vspace{10mm}
\section{Introduction}\label{intro}
It was Henri Poincar\'e who worked out the mathematics of Lorentz
transformations before Einstein and Minkowski, and the Poincar\'e group
is the underlying language for special relativity.  In order to analyze
the polarization of light, Poincar\'e also constructed a graphic
illustration known as the Poincar\'e sphere~\cite{azzam77,born80,bros98}.
\par
It is of interest to see whether the Poincar\'e sphere can also speak
the language of special relativity. In that case, we can study
the physics of relativity in terms of what we observe in optical
laboratories.  For that purpose, we note first that the Lorentz group
starts as a group of four-by-four matrices, while the Poincar\'e sphere
is based on the two-by-two matrix consisting of four Stokes parameters.
Thus, it is essential to find a two-by-two representation of the Lorentz
group.  Fortunately, this representation exists in the
literature~\cite{naimark54,naimark64}, and we shall use it in this paper.
\par
As for the problems in relativity, we shall discuss here Wigner's
little groups dictating the internal space-time symmetries of
relativistic particles~\cite{kiwi90jmp}.  In his original paper of
1939~\cite{wig39}, Wigner considered the subgroups of the Lorentz
group whose transformations leave the four-momentum of a given particle
invariant.  While this problem has been extensively discussed in the
literature, we propose here to study it using Naimark's two-by-two
representation of the Lorentz group~\cite{naimark54,naimark64}.
\par
This two-by-two representation is useful for communicating with the
symmetries of the Poincar\'e sphere based on the four Stokes parameters,
which can take the form of two-by-two matrices.  We shall prove here
that the Poincar\'e sphere shares the same symmetry property as that of
the Lorentz group, particularly in approaching Wigner's little groups.
By doing this, we can study the Lorentz symmetries of elementary particles
from what we observe in optical laboratories.
\par
The present paper starts from an unpublished note based on an invited
paper presented by one of the authors (YSK) at the {\em Fedorov Memorial
Symposium: Spins and Photonic Beams at Interface} held in Minsk
(2011)~\cite{ysk12minsk}.  To this, we have added a detailed discussion
of how the decoherence mechanism in polarization optics is mathematically
equivalent to a massless particle gaining mass to become a massive particle.
We are particularly interested how the variation of mass can be accommodated
in the study of internal space-time symmetries.
\par
In Sec.~\ref{poincw}, we define the symmetry problem we propose to study
in this paper.  We are interested in the subgroups of the Lorentz group
whose transformations leave the four-momentum of a given particle
invariant.  This is an old problem and has been repeatedly discussed in
the literature~\cite{kiwi90jmp,wig39,knp86}.  In this paper, we discuss
this problem using the two-by-two formulation of the Lorentz group.
This two-by-two language is directly applicable to polarization optics
and the Poincar\'e sphere.
\par

While Wigner formulated his little groups for particles in their given
Lorentz frames, we give a formalism applicable to all Lorentz frames.
In his 1939 paper, Wigner pointed out that his little groups are different
for massive, massless, and imaginary-particles.
In Sec.~\ref{complete}, we discuss the possibility of deriving the
symmetry properties for massive and imaginary-mass particles from that
of the massless particle.
\par
In Sec.~\ref{jones}, we assemble the variables in polarization optics,
and define the matrix operators corresponding to transformations
applicable to those variables.  We write the Stokes parameters in the
form of a two-by-two matrix.  The Poincar\'e sphere can be constructed
from this two-by-two Stokes matrix.
In Sec.~\ref{poincs}, we note that there can be two radii for the
Poincar\'e sphere.  Poincar\'e's original sphere has one fixed radius,
but this radius can change depending on the degree of coherence.
Based on what we studied in Sec.~\ref{complete}, we can associate
this change of the radius to the change in mass of the particle.

\section{Poincar\'e Group and Wigner's Little Groups}\label{poincw}

Poincar\'e formulated the group theory of Lorentz transformations
applicable to the four-dimensional space consisting of three space
coordinates and one time variable.   There are six generators for
this group consisting of three rotation and three boost generators.

\par
In addition, Poincar\'e considered translations applicable to those
four space-time variables, with four generators.  If we add these
four generators to the six generators for the homogenous Lorentz group,
the result is the inhomogeneous Lorentz group~\cite{wig39} with ten
generators.  This larger group is called the Poincar\'e group in the
literature.

\par
The four translation generators produce space-time four-vectors
consisting of the energy and momentum.  Thus, within the framework of
the Poincar\'e group, we can consider the subgroup of the Lorentz group
for a fixed value of momentum~\cite{wig39}.  This subgroup defines the
internal space-time symmetry of the particle.  Let us consider a particle
at rest.  Its momentum consists of its mass as its time-like variable
and zero for the three momentum components.
\begin{equation}\label{fm01}
     (m, 0, 0, 0).
\end{equation}
For convenience, we use the four-vector convention $(t, z, x, y)$
and $\left(E, p_z, p_x, p_y\right)$.
\par
This four-momentum of Eq.(\ref{fm01}) is invariant under three-dimensional
rotations applicable only to the $z, x, y$ coordinates.  The dynamical
variable associated with this rotational degree of freedom is called the
spin of the particle.
\par
We are then interested in what happens when the particle moves with a
non-zero momentum.  If it moves along the $z$ direction, the four-momentum
takes the value
\begin{equation}\label{fm02}
    m (\cosh\eta, \sinh\eta, 0, 0) ,
\end{equation}
which means
\begin{equation}\label{fm03}
   p_0 = m(\cosh\eta), \quad p_z = m(\sinh\eta), \quad
e^{\eta} = \sqrt{\frac{p_0 + p_z}{p_0 - p_z}} .
\end{equation}
Accordingly, the little group consists of Lorentz-boosted rotation
matrices.  This aspect of the little group has been discussed in the
literature~\cite{kiwi90jmp,knp86}.  The question then is whether we
could carry out the same logic using two-by-two matrices
\par
Of particular interest is what happens when the transformation parameter
$\eta$ becomes very large, and the four-momentum becomes that of a
massless particle.  This problem has also been discussed in the literature
within the framework of four-dimensional Minkowski space.  The $\eta$
parameter becomes large when the momentum becomes large, but it can also
become large when the mass becomes very small.  The two-by-two formulation
allows us to study these two cases separately, as we will do in
Sec.~\ref{complete}.
\par
If the particle has an imaginary mass, it moves faster than light and
is not observable.  Yet, particles of this kind play important roles
in Feynman diagrams, and their space-time symmetry should also be
studied.  In his original paper~\cite{wig39}, Wigner studied the little
group as the subgroup of the Lorentz group whose transformations leave
invariant the four-momentum of the form
\begin{equation}
 (0, k, 0, 0) .
\end{equation}
Wigner observed that this four-momentum remains invariant under the
Lorentz boost along the $x$ or $y$ direction.
\par
If we boost this four-momentum along the $z$ direction,
the four-momentum becomes
\begin{equation}\label{superlum03}
k(\sinh\eta, \cosh\eta, 0, 0) ,
\end{equation}
with
\begin{equation} \label{superlum05}
e^{\eta} = \sqrt{\frac{p_0 + p_z}{p_z - p_0}} .
\end{equation}
The two-by-two formalism also allows us to study this problem.
\par
In Subsec.~\ref{twobytwo}, we shall present the two-by-two representation
of the Lorentz group.  In Subsec.~\ref{wigner}, we shall present Wigner's
little groups in this two-by-two representation.  While Wigner's analysis
was based on particles in their fixed Lorentz frames, we are
interested in what happens when they start moving.  We shall deal with
this problem in Sec.~\ref{complete}.

\subsection{Two-by-two Representation of the Lorentz groups}\label{twobytwo}
The Lorentz group starts with a group of four-by-four matrices
performing Lorentz transformations on the Minkowskian vector
space of $(t, z, x, y),$ leaving the quantity
\begin{equation}\label{4vec02}
t^2 - z^2 - x^2 - y^2
\end{equation}
invariant.  It is possible to perform this transformation using two-by-two
representations~\cite{naimark54,naimark64}.  This mathematical aspect is
known as $SL(2,c)$, the universal covering group for the Lorentz group.

\par
In this two-by-two representation, we write the four-vector as a matrix
\begin{equation}
X = \pmatrix{t + z  &  x - iy \cr x + iy & t - z} .
\end{equation}
Then its determinant is precisely the quantity given in Eq.(\ref{4vec02}).
Thus the Lorentz transformation on this matrix is a determinant-preserving
transformation.  Let us consider the transformation matrix as
\begin{equation}\label{g22}
 G = \pmatrix{\alpha & \beta \cr \gamma & \delta}, \qquad G^{\dagger} =
  \pmatrix{\alpha^* & \gamma^* \cr \beta^* & \delta^*} ,
\end{equation}
with
\begin{equation}
    \det{(G)} = 1.
\end{equation}
The $G$ matrix starts with four complex numbers.  Due to the above
condition on its determinant, it has six independent parameters.  The
group of these $G$ matrices is known to be locally isomorphic to the
group of four-by-four matrices performing Lorentz transformations on
the four-vector $(t, z, x, y)$.  In other words, for each $G$ matrix
there is a corresponding four-by-four Lorentz-transform matrix, as is
illustrated in the Appendix.
\par

The matrix $G$ is not a unitary matrix, because its Hermitian conjugate
is not always its inverse.   The group can have a unitary subgroup
called $SU(2)$ performing rotations on electron spins.  As far as we
can see, this $G$-matrix formalism was first presented by Naimark in
1954~\cite{naimark54}.  Thus, we call this formalism the Naimark
representation of the Lorentz group.  We shall see first that this
representation is convenient for studying space-time symmetries of
particles.  We shall then note that this Naimark representation is the
natural language for the Stokes parameters in polarization optics.

\par
With this point in mind, we can now consider the transformation
\begin{equation}\label{naim}
X' = G X G^{\dagger} .
\end{equation}
Since $G$ is not a unitary matrix, it is not a unitary transformation.
In order to tell this difference, we call this the ``Naimark
transformation.''  This expression can be written explicitly as
\begin{equation}\label{lt01}
\pmatrix{t' + z' & x' - iy' \cr x + iy & t' - z'}
 = \pmatrix{\alpha & \beta \cr \gamma & \delta}
  \pmatrix{t + z & x - iy \cr x + iy & t - z}
  \pmatrix{\alpha^* & \gamma^* \cr \beta^* & \delta^*} ,
\end{equation}
\par
For this transformation, we have to deal with four complex numbers.
However, for all practical purposes, we may work with two Hermitian matrices
\begin{equation}\label{herm11}
Z(\delta) = \pmatrix{e^{i\delta/2} & 0 \cr 0 & e^{-i\delta/2}}, \qquad
R(\delta) = \pmatrix{\cos(\theta/2)  & -\sin(\theta/2) \cr
     \sin(\theta/2) & \cos(\theta/2)} ,
\end{equation}
and two symmetric matrices
\begin{equation}\label{symm11}
B(\eta) = \pmatrix{e^{\eta/2} & 0 \cr 0 & e^{-\eta/2}}, \qquad
S(\lambda) = \pmatrix{\cosh(\lambda/2)  & \sinh(\lambda/2) \cr
     \sinh(\lambda/2) & \cosh(\lambda/2)},
\end{equation}
whose Hermitian conjugates are not their inverses.  The two Hermitian
matrices in Eq.(\ref{herm11}) lead to rotations around the $z$ and $y$
axes respectively.  The symmetric matrices in Eq.(\ref{symm11}) perform
Lorentz boosts along the $z$ and $x$ directions respectively.

\par

Repeated applications of these four matrices will lead to the most general
form of the $G$ matrix of Eq.(\ref{g22}) with six independent parameters.
For each two-by-two Naimark transformation, there is a four-by-four matrix
performing the corresponding Lorentz transformation on the four-component
four-vector.  In the Appendix, the four-by-four equivalents are given
for the matrices of Eq.(\ref{herm11})
and Eq.(\ref{symm11}).

\par

It was Einstein who defined the energy-momentum four-vector, and showed
that it also has the same Lorentz-transformation law as the space-time
four-vector.  We write the energy-momentum four-vector as
\begin{equation}\label{mom11}
P = \pmatrix{E + p_z & p_x - ip_y \cr p_x + ip_y & E - p_z} ,
\end{equation}
with
\begin{equation}
\det{(P)} = E^2 - p_x^2 - p_y^2 - p_z^2,
\end{equation}
which means
\begin{equation}\label{mass}
\det{(P)} = m^2,
\end{equation}
where $m$ is the particle mass.
\par
Now Einstein's transformation law can be written as
 \begin{equation}
 P' = G P G^{\dagger} ,
 \end{equation}
or explicitly
\begin{equation}\label{lt03}
\pmatrix{E' + p_z' & p_x' - ip_y' \cr p'_x + ip'_y & E' - p'_z}
 = \pmatrix{\alpha & \beta \cr \gamma & \delta}
  \pmatrix{E + p_z & p_x - ip_y \cr p_x + ip_y & E - p_z}
  \pmatrix{\alpha^* & \gamma^* \cr \beta^* & \delta^*} .
\end{equation}

\subsection{Wigner's Little Groups}\label{wigner}
Later in 1939~\cite{wig39}, Wigner was interested in constructing
subgroups of the Lorentz group whose transformations leave a given
four-momentum invariant.  He called these subsets ``little groups.''
Thus, Wigner's little group consists of two-by-two matrices satisfying
\begin{equation}\label{wigcon}
P = W P W^{\dagger} .
\end{equation}
This two-by-two $W$ matrix is not an identity matrix, but tells
about the internal space-time symmetry of a particle with a given
energy-momentum four-vector.  This aspect was not known when Einstein
formulated his special relativity in 1905.  The internal space-time
symmetry was not an issue at that time.

\par

If its determinant is a positive number, the $P$ matrix can be
brought to a form proportional to
\begin{equation}\label{massive}
         P = \pmatrix{1 & 0 \cr 0 & 1},
\end{equation}
corresponding to a massive particle at rest.
\par
If the determinant is negative, it can be brought to a form
proportional to
\begin{equation}\label{superlum}
         P = \pmatrix{1 & 0 \cr 0 & -1} ,
\end{equation}
corresponding to an imaginary-mass particle moving faster than light
along the $z$ direction, with its vanishing energy component.

\par

\begin{table}[h]
\caption{Wigner's Little Groups.  The little groups are the subgroups of
the Lorentz group whose transformations leave the four-momentum of a given
particle invariant.  Thus, the little groups define the internal space-time
symmetries of particles.  The four-momentum remains invariant under the
rotation around it. In addition, the four-momentum remains invariant under
the following transformations. These  transformations are different for
massive, massless, and imaginary-mass particles.}\label{tab11}
\vspace{2mm}
\begin{center}
\begin{tabular}{llclc}
\hline
\hline \\[0.5ex]
 Particle mass &{}&  Four-momentum  &\hspace{7mm} &  Transform matrices \\[1.0ex]
\hline\\
Massive  &{}& $\pmatrix{1 & 0 \cr 0 & 1}$
&{}&
$\pmatrix{\cos(\theta/2) & -\sin(\theta/2)\cr \sin(\theta/2) & \cos(\theta/2)}$
\\[4ex]
Massless  &{}&
$\pmatrix{1 & 0 \cr 0 & 0}$
&{}& $\pmatrix{1 & \gamma \cr 0 & 1}$
\\[4ex]
Imaginary mass &{}&
$\pmatrix{1 & 0\cr 0 & -1}$
&{}&  $\pmatrix{\cosh(\lambda/2) & \sinh(\lambda/2) \cr \sinh(\lambda/2) & \cosh(\lambda/2)}$
\\[4ex]
\hline
\hline\\[-0.8ex]
\end{tabular}
\end{center}
\end{table}

If the determinant is zero, we may write $P$ as
\begin{equation}\label{mzero}
         P = \pmatrix{1 & 0 \cr 0 & 0} ,
\end{equation}
which is proportional to the four-momentum matrix for a massless particle
moving along the $z$ direction.
\par
For all three of the above cases, the matrix of the form
\begin{equation}\label{22rotz}
Z(\delta) = \pmatrix{e^{i\delta/2} & 0 \cr 0 & e^{-i\delta/2}}
\end{equation}
will satisfy the Wigner condition of Eq.(\ref{wigcon}).  This matrix
corresponds to rotations around the $z$ axis, as is shown in the Appendix.

\par
For the massive particle with the four-momentum of Eq.(\ref{massive}),
the Naimark transformations with the rotation matrix of the form
\begin{equation}\label{rot55}
R(\theta) = \pmatrix{\cos(\theta/2) & -\sin(\theta/2) \cr
\sin(\theta/2) & \cos(\theta/2)} ,
\end{equation}
also leaves the $P$ matrix of Eq.(\ref{massive}) invariant.  Together
with the $Z(\delta)$ matrix, this rotation matrix leads to the subgroup
consisting of the unitary subset of the $G$ matrices.  The unitary subset
of $G$ is $SU(2)$ corresponding to the three-dimensional rotation group
dictating the spin of the particle~\cite{knp86}.
\par
For the massless case, the transformations with the triangular matrix
of the form
\begin{equation}\label{trian}
\pmatrix{1 & \gamma \cr 0 & 1}
\end{equation}
leave the momentum matrix of Eq.(\ref{mzero}) invariant.  The physics
of this matrix has a stormy history, and the variable $\gamma$
leads to gauge transformation applicable to massless
particles~\cite{kiwi90jmp,hks82}.
\par
For a particle with its imaginary mass, the $W$ matrix of the form
\begin{equation}
S(\lambda) = \pmatrix{\cosh(\lambda/2) & \sinh(\lambda/2) \cr
 \sinh(\lambda/2) & \cosh(\lambda/2)}
\end{equation}
will leave the four-momentum of Eq.(\ref{superlum}) invariant.  This
unobservable particle does not appear to have observable internal
space-time degrees of freedom.

\par

Table~\ref{tab11} summarizes the transformation matrices for Wigner's
subgroups for massive, massless, and imaginary-mass particles.  Of
course, it is a challenging problem to have one expression for all
those three cases, and this problem has been addressed in the
literature~\cite{bk10jmo}.

\section{Lorentz Completion of Wigner's Little Groups}\label{complete}
In his original paper~\cite{wig39}, Wigner worked out his little groups
for specific Lorentz frames.  For the massive particle, he constructed
his little group in the frame where the particle is at rest.  For
the imaginary-mass particle, the energy-component of his frame is zero.
\par
For the massless particle, it moves along the $z$ direction with a
nonzero momentum.  There are no specific frames particularly convenient
for us.  Thus, the specific frame can be chosen for an arbitrary value
of the momentum, and the triangular matrix of Eq.(\ref{trian}) should
remain invariant under Lorentz boosts along the $z$ direction.

\par
For the massive particle, let us Lorentz-boost the four-momentum matrix
of Eq.(\ref{massive}) by performing a Naimark transformation:
\begin{equation}
\pmatrix{e^{\eta/2}& 0 \cr 0 & e^{-\eta/2}} \pmatrix{1 & 0 \cr 0 & 1}
\pmatrix{e^{\eta/2}& 0 \cr 0 & e^{-\eta/2}} ,
\end{equation}
which leads to
\begin{equation}\label{mat01}
\pmatrix{e^{\eta} & 0 \cr 0 & e^{-\eta}} .
\end{equation}
This resulting matrix corresponds to the Lorentz-boosted four-momentum
given in Eq.(\ref{fm02}).  For simplicity, we let $m = 1$ hereafter
in this paper.   The Lorentz transformation applicable to the
four-momentum matrix is not a similarity transformation, but it is a
Naimark transformation as defined in Eq.(\ref{naim}).
\par
On the other hand, the rotation matrix of Eq.(\ref{rot55}) is
Lorentz-boosted as a similarity transformation:
\begin{equation}\label{mat09}
\pmatrix{e^{\eta/2} & 0 \cr 0 & e^{-\eta/2}}
\pmatrix{\cos(\theta/2) & -\sin(\theta/2) \cr
       \sin(\theta/2) & \cos(\theta/2)}
\pmatrix{e^{-\eta/2} & 0 \cr 0 & e^{\eta/2}},
\end{equation}
and it becomes
\begin{equation} \label{mat11}
\pmatrix{\cos(\theta/2) & - e^{\eta}\sin(\theta/2) \cr
e^{-\eta}\sin(\theta/2) & \cos(\theta/2)} .
\end{equation}
If we perform the Naimark transformation of the four-momentum matrix
of Eq.(\ref{mat01}) with this Lorentz-boosted rotation matrix:
\begin{equation}
\pmatrix{\cos(\theta/2) & - e^{\eta}\sin(\theta/2)
   \cr e^{-\eta/2} \sin(\theta/2) & \cos(\theta/2)}
\pmatrix{e^{\eta} & 0 \cr 0 & e^{-\eta}}
\pmatrix{\cos(\theta/2) & e^{\eta}\sin(\theta/2)
    \cr -e^{-\eta} \sin(\theta/2) & \cos(\theta/2)},
\end{equation}
the result is the four-momentum matrix of Eq.(\ref{mat01}).
This means that the Lorentz-boosted rotation matrix of Eq.(\ref{mat11})
represents the little group whose transformations leave the four-momentum
matrix of Eq.(\ref{mat01}) invariant.

\par
For the imaginary-mass case, the Lorentz boosted four-momentum matrix
becomes
\begin{equation}\label{mat02}
\pmatrix{e^{\eta} & 0 \cr 0 & -e^{-\eta}} .
\end{equation}
The little group matrix is
\begin{equation}\label{mat03}
\pmatrix{\cosh(\lambda/2) & e^{\eta}\sinh(\lambda/2)
   \cr e^{-\eta} \sinh(\lambda/2) & \cosh(\lambda/2)} ,
\end{equation}
where $\eta$ is given in Eq.(\ref{superlum05}).
\par
For the massless case, if we boost the four-momentum matrix of
Eq.(\ref{mzero}), the result is
\begin{equation}
e^{\eta} \pmatrix{1 & 0 \cr 0 & 0} .
\end{equation}
Here $\eta$ parameter is an independent variable and cannot be
defined in terms of the momentum or energy.

\par
The remaining problem is to see whether the massive and imaginary-mass
cases collapse to the massless case in the large $\eta$ limit.  This
variable becomes large when the momentum becomes large or the mass
becomes small.  We shall discuss these two cases separately.

\subsection{Large-momentum limit}

While Wigner defined his little group for the massive particle in its
rest frame in his original paper~\cite{wig39}, the little group
represented by Eq.(\ref{mat11}) is applicable to the moving particle
whose four-momentum is given in Eq.(\ref{mat01}).  This matrix can
also be written as
\begin{equation}
e^{\eta}\pmatrix{1 & 0 \cr 0 & e^{-2\eta} } .
\end{equation}
In the limit of large $\eta$, we can change the above expression into
\begin{equation}\label{mat55}
e^{\eta}\pmatrix{1 & 0 \cr 0 & 0} .
\end{equation}
This process is continuous, but not necessarily analytic~\cite{bk10jmo}.
After making this transition, we can come back to the original frame to
obtain the four momentum matrix of Eq.(\ref{mzero}).

\par
The remaining problem is the Lorentz-boosted rotation matrix of
Eq.(\ref{mat11}).  If this matrix is going to remain finite as $\eta$
approaches infinity, the upper-right element should be finite for large
values of $\eta$.  Let it be $\gamma$.  Then
\begin{equation}
-e^{\eta}\sin(\theta/2) = \gamma.
\end{equation}
This means that angle $\theta$ has to become zero.  As a consequence, the
little group matrix of Eq.(\ref{mat11}) become the triangular matrix given
in Eq.(\ref{trian}) for massless particles.
\par

Imaginary-mass particles move faster than light, and they are not
observable.  On the other hand, the mathematics applicable to Wigner's
little group for this particle has been useful in the two-by-two beam
transfer matrix in ray and polarization optics~\cite{bk13mop}.
\par
Let us go back to the four-momentum matrix of Eq.(\ref{superlum}).
If we boost this matrix, it becomes
\begin{equation}\label{mat33}
\pmatrix{e^{\eta} & 0 \cr 0 & -e^{-\eta}} ,
\end{equation}
which can be written as
\begin{equation}
e^{\eta}\pmatrix{1 & 0 \cr 0 & -e^{-2\eta}}.
\end{equation}
This matrix can be changed to the form Eq.(\ref{mat55}) in the limit of
large $\eta$.
\par
Indeed, the little groups for massive, massless, and imaginary cases coincide
in the large-$\eta$ limit.  Thus, it is possible to jump from one little
group to another, and it is a continuous process but not necessarily
analytic~\cite{bk13mop}.
\par
The $\eta$ parameter can become large as the momentum becomes large or the
mass becomes small.  In this subsection, we considered the case for large
momentum.  However, it is of interest to see the limiting process when
the mass becomes small, especially in view of the fact that neutrinos have
small masses.

\subsection{Small-mass limit}
Let us start with a massive particle with fixed energy $E$.
Then, $p_0 = E$, and $p_z = E\cos\chi$.  The  four-momentum matrix
is
\begin{equation}\label{mat31}
E \pmatrix{1 + \cos\chi & 0 \cr 0 & 1 - \cos\chi} .
\end{equation}
The determinant of this matrix is $E^2 (\sin\chi)^2$.   In the regime of
the Lorentz group, this is the $(mass)^2$, and is a Lorentz-invariant
quantity.  There are no Lorentz transformations which change the angle $\chi$.
Thus, with this extra variable, it is possible to study the little groups
for variable masses, including the small-mass limit and the zero-mass case.
\par
If $\chi = 0$, the matrix of Eq.(\ref{mat31}) becomes that of the four-momentum
matrix for a  massless particle.  As it becomes a positive small number,
the matrix of Eq.(\ref{mat31}) can be written as
\begin{equation}\label{mat88}
E(\sin\chi) \pmatrix{e^{\eta} & 0 \cr 0 & e^{-\eta}} ,
\end{equation}
with
\begin{equation}
e^\eta = \sqrt{\frac{1 + \cos\chi}{1 - \cos\chi}} .
\end{equation}
Here again, the determinant of Eq.(\ref{mat88}) is $E^2 (\sin\chi)^2$.
With this matrix, we can construct Wigner's little group for each value
of the angle $\chi.$  If $\chi$ is not zero, even if it is very small,
the little group is $O(3)$-like as in the case of all massive particles.
As the angle $\chi$ varies continuously from zero to $90^o$, the mass
increases from zero to its maximum value.
\par
It is important to note that the little groups are different for the
small-mass limit and for the zero-mass case.  In this section, we
studied the internal space-time symmetries dictated by Wigner's little
groups, and we are able to present their Lorentz-covariant picture in
Table~\ref{tab22}.

\begin{table}
\caption{Covariance of the energy-momentum relation, and covariance
of the internal space-time symmetry groups.  The $\gamma$ parameter
for the massless case has been studied in earlier papers in the
four-by-four matrix formulation~\cite{kiwi90jmp}.  It corresponds to
a gauge transformation.  Among the three spin components, $S_3$
is along the direction of the momentum and remains invariant.
It is called the "helicity." }\label{tab22}
\vspace{5mm}
\begin{center}
\begin{tabular}{ccccc}

\hline\hline \\[0.5ex]
Massive, Slow &\hspace{7mm} & COVARIANCE &\hspace{10mm}& Massless, Fast \\[2mm]
\hline\\
$E = p^{2}/2m$ &{}& Einstein's $E = mc^{2}$ &{}& $E = cp$ \\[4mm]
\hline \\
$S_{3}$ &{}& {}  &{}&   Helicity \\ [-1mm]
{} &{}& Wigner's Little Group &{}& {} \\[-1mm]
$S_{1}, S_{2}$ &{}& {} &{}& Gauge Transformation \\[4mm]
\hline\hline\\[-0.8ex]
\end{tabular}
\end{center}
\end{table}

\section{Jones Vectors and Stokes Parameters}\label{jones}

In studying polarized light propagating along the $z$ direction, the
traditional approach is to consider the $x$ and $y$ components of
the electric fields.  Their amplitude ratio and the phase difference
determine the state of polarization.  Thus, we can change the polarization
either by adjusting the amplitudes, by changing the relative phase,
or both.  For convenience, we call the optical device which changes
amplitudes an ``attenuator'' and the device which changes the relative
phase a ``phase shifter.''
\par
The traditional language for this two-component light is the Jones-vector
formalism which is discussed in standard optics textbooks~\cite{saleh07}.
In this formalism, the above two components are combined into one column
matrix with the exponential form for the sinusoidal function
\begin{equation}\label{jvec11}
\pmatrix{\psi_1(z,t) \cr \psi_2(z,t)} =
\pmatrix{a \exp{\left\{i(kz - \omega t + \phi_{1})\right\}}  \cr
b \exp{\left\{i(kz - \omega t + \phi_{2})\right\}}} .
\end{equation}
This column matrix is called the Jones vector.
\par

When the beam goes through a medium with different values of indexes of
refraction for the $x$ and $y$ directions, we have to apply the matrix
\begin{equation}\label{phase3}
\pmatrix{e^{i\delta_{1}} & 0 \cr 0 & e^{i\delta_{2}}}
= e^{i(\delta_{1} + \delta_{2})/2}
\pmatrix{e^{-i\delta/2} & 0 \cr 0 & e^{i\delta/2}} ,
\end{equation}
with $\delta = \delta_{1} - \delta_{2}$ .
In measurement processes, the overall phase factor
$e^{i(\delta_{1} + \delta_{2})/2}$
cannot be detected, and can therefore be deleted.  The polarization
effect of the filter is solely determined by the matrix
\begin{equation}\label{shif11}
Z(\delta) = \pmatrix{e^{i\delta/2} & 0 \cr 0 & e^{-i\delta/2}} ,
\end{equation}
which leads to a phase difference of $\delta$ between the $x$ and $y$
components.  The form of this matrix is given in Eq.(\ref{herm11}), which
serves as the rotation around the $z$ axis in the Minkowski space and
time.
\par
Also along the $x$ and $y$ directions, the attenuation coefficients
could be different.  This will lead to the matrix~\cite{hkn97josa}
\begin{equation}\label{atten}
\pmatrix{e^{-\eta_{1}} & 0 \cr 0 & e^{-\eta_{2}}}
   = e^{-(\eta_{1} + \eta_{2})/2} \pmatrix{e^{\eta/2} & 0 \cr 0 & e^{-\eta/2}}
\end{equation}
with $\eta = \eta_{2} - \eta_{1}$ .
If $\eta_1 = 0$ and $\eta_2 = \infty$, the above matrix becomes
\begin{equation}\label{polar}
\pmatrix{1 & 0 \cr 0 & 0} ,
\end{equation}
which eliminates the $y$ component.  This matrix is known as a polarizer
in the textbooks~\cite{saleh07}, and is a special case of the attenuation
matrix of Eq.(\ref{atten}).

\par
This attenuation matrix tells us that the electric fields are attenuated
at two different rates.  The exponential factor $e^{-(\eta_{1} + \eta_{2})/2}$
reduces both components at the same rate and does not affect the state of
polarization.  The effect of polarization is solely determined by the
squeeze matrix~\cite{hkn97josa}
\begin{equation}\label{sq11}
B(\eta) = \pmatrix{e^{\eta/2} & 0 \cr 0 & e^{-\eta/2}} .
\end{equation}
This diagonal matrix is given in Eq.(\ref{symm11}).  In the language of
space-time symmetries, this matrix performs a Lorentz boost along the
$z$ direction.

\par
The polarization axes are not always the $x$ and $y$ axes.
For this reason, we need the rotation matrix
\begin{equation}\label{rot11}
R(\theta) = \pmatrix{\cos(\theta/2) & -\sin(\theta/2)
\cr \sin(\theta/2) & \cos(\theta/2)} ,
\end{equation}
which, according to Eq.(\ref{herm11}), corresponds to the rotation around
the $y$ axis in the space-time symmetry.
\par
Among the rotation angles, the angle of $45^o$ plays an important role in
polarization optics.  Indeed, if we rotate the squeeze matrix of Eq.(\ref{sq11})
by $45^o$, we end up with the squeeze matrix
\begin{equation}\label{sq22}
R(\theta) = \pmatrix{\cosh(\lambda/2) & \sinh(\lambda/2)
\cr \sinh(\lambda/2) & \cosh(\lambda/2)} ,
\end{equation}
which is also given in Eq.(\ref{symm11}).  In the language of space-time physics,
this matrix leads to a Lorentz boost along the $x$ axis.
\par
Indeed, the $G$ matrix of Eq.(\ref{g22}) is the most general form of the
transformation matrix applicable to the Jones vector.  Each of the above four
matrices plays its important role in special relativity, as we discussed in
Sec.~\ref{poincw}.  Their respective roles in optics and particle physics are
given in Table~\ref{tab33}.

\par

However, the Jones vector alone cannot tell us whether the two components are
coherent with each other.  In order to address this important degree of freedom,
we use the coherency matrix~\cite{azzam77,born80}
\begin{equation}\label{cocy11}
C = \pmatrix{S_{11} & S_{12} \cr S_{21} & S_{22}},
\end{equation}
with
\begin{equation}
<\psi_{i}^* \psi_{j}> = \frac{1}{T} \int_{0}^{T}\psi_{i}^* (t + \tau) \psi_{j}(t) dt,
\end{equation}
where $T$, for a sufficiently long time interval, is much larger than $\tau$.
Then, those four elements become~\cite{hkn97}
\begin{eqnarray}
&{}& S_{11} = <\psi_{1}^{*}\psi_{1}> =a^2  , \qquad
S_{12} = <\psi_{1}^{*}\psi_{2}> = ab~e^{-(\sigma +i\delta)} , \nonumber \\[1ex]
&{}& S_{21} = <\psi_{2}^{*}\psi_{1}> = ab~e^{-(\sigma -i\delta)} ,  \qquad
S_{22} = <\psi_{2}^{*}\psi_{2}>  = b^2 .
\end{eqnarray}
The diagonal elements are the absolute values of $\psi_1$ and $\psi_2$
respectively.  The off-diagonal elements could be smaller than the product
of $\psi_1$ and $\psi_2$, if the two beams are not completely coherent.
The $\sigma$ parameter specifies the degree of coherency.
\par
This coherency matrix is not always real but it is Hermitian.  Thus it can be
diagonalized by a unitary transformation.  If this matrix is normalized so
that its trace is one, it becomes a density matrix~\cite{fey72,hkn99ajp}.
\par

\begin{table}
\caption{Polarization optics and special relativity sharing the same mathematics.
Each matrix has its clear role in both optics and relativity.  The determinant
of the Stokes or the four-momentum matrix remains invariant under Lorentz
transformations.  It is interesting to note that the decoherence parameter
(least fundamental) in optics corresponds to the mass (most fundamental) in
particle physics.}\label{tab33}
\vspace{2mm}
\begin{center}
\begin{tabular}{llcll}
\hline
\hline \\[0.5ex]
 Polarization Optics &\hspace{10mm}& Transformation Matrix  &\hspace{10mm} &
 Particle Symmetry \\[1.0ex]
\hline \\
Phase shift $\delta$  &{}&
$\pmatrix{e^{\delta/2} & 0\cr 0 & e^{-i\delta/2}}$
&{}&  Rotation around $z$.
\\[4ex]
Rotation around $z$  &{}&
$\pmatrix{\cos(\theta/2) & -\sin(\theta/2)\cr \sin(\theta/2) & \cos(\theta/2)}$
&{}&  Rotation around  $y$.
\\[4ex]
Squeeze along $x$ and $y$  &{}&
$\pmatrix{e^{\eta/2} & 0\cr 0 & e^{-\eta/2}}$
&{}&  Boost along $z$.
\\[4ex]
Squeeze along $45^o$  &{}&
$\pmatrix{\cosh(\lambda/2) & \sinh(\lambda/2)\cr \sinh(\lambda/2)
                & \cosh(\lambda/2)}$
&{}&   Boost along $x$.
\\[4ex]
$(ab)^{2} \sin^2\chi$  &{}& Determinant &{}&  (mass)$^2$
\\[4ex]
\hline
\hline\\[-0.8ex]
\end{tabular}
\end{center}
\end{table}

If we start with the Jones vector of the form of Eq.(\ref{jvec11}), the coherency
matrix becomes
\begin{equation}\label{cocy22}
C = \pmatrix{a^2 & ab~e^{-(\sigma + i\delta)} \cr
ab~e^{-(\sigma - i\delta)} & b^2} .
\end{equation}
We are interested in the symmetry properties of this matrix.  Since the
transformation matrix applicable to the Jones vector is the two-by-two
representation of the Lorentz group, we are particularly interested in the
transformation matrices applicable to this coherency matrix.
\par
The trace and the determinant of the above coherency matrix
are
\begin{eqnarray}
&{}& \det(C) = (ab)^2 \left(1 - e^{-2\sigma}\right), \nonumber \\[2ex]
&{}& \mbox{tr}(C) = a^2 + b^2 .
\end{eqnarray}
Since $e^{-\sigma}$ is always smaller than one, we can introduce
an angle $\chi$ defined as
\begin{equation}
\cos\chi = e^{-\sigma} ,
\end{equation}
and call it the ``decoherence angle.''  If $\chi = 0$, the decoherence is
minimum, and it becomes maximum when $\chi = 90^o$.  We can then write the
coherency matrix of Eq.(\ref{cocy22}) as
\begin{equation}\label{cocy22b}
C = \pmatrix{a^2 & ab(\cos\chi)e^{-i\delta} \cr
ab(\cos\chi)e^{i\delta} & b^2}.
\end{equation}

\par
The degree of polarization is defined as~\cite{saleh07}
\begin{equation}\label{pdegree}
f = \sqrt{ 1 - \frac{4~\det(C)}{(\mbox{tr}(C))^2}} =
        \sqrt{1 - \frac{4(ab)^2\sin^2\chi}{(a^2 + b^2)^2}} .
\end{equation}
This degree is one if $\chi = 0$.  When $\chi = 90^o$, it becomes
\begin{equation}
  \frac{a^2 - b^2}{a^2 + b^2} ,
\end{equation}
Without loss of generality, we can assume that $a$ is greater than $b$.
If they are equal, this minimum degree of polarization is zero.
\par

Under the influence of the Naimark transformation given in Eq.(\ref{naim}),
this coherency matrix is transformed as
\begin{eqnarray}\label{trans22}
&{}& C' = G~C~G^{\dagger} =
\pmatrix{S'_{11} & S'_{12} \cr S'_{21} & S'_{22}} \nonumber \\[2ex]
&{}&\hspace{5ex} = \pmatrix{\alpha & \beta \cr \gamma & \delta}
\pmatrix{S_{11} & S_{12} \cr S_{21} & S_{22}}
\pmatrix{\alpha^{*} & \gamma^{*} \cr \beta^{*} & \delta^{*}} .
\end{eqnarray}
It is more convenient to make the following linear combinations.
\begin{eqnarray}\label{stokes11}
&{}& S_{0} = \frac{S_{11} + S_{22}}{2},  \qquad
    S_{3} = \frac{S_{11} - S_{22}}{2},    \nonumber \\[2ex]
&{}& S_{1} = \frac{S_{12} + S_{21}}{2}, \qquad
S_{2} = \frac{S_{12} - S_{21}}{2i}.
\end{eqnarray}
These four parameters are called Stokes parameters, and four-by-four
transformations applicable to these parameters are widely known as
Mueller matrices~\cite{azzam77,bros98}.
However, if the Naimark transformation given in Eq.(\ref{trans22}) is
translated into the four-by-four Lorentz transformations according to the
correspondence given in the Appendix,  the Mueller matrices constitute
a representation of the Lorentz group.

\par

Another interesting aspect of the two-by-two matrix formalism is that
the coherency matrix can be formulated in terms of
quarternions~\cite{pellat91,dlugu09,tudor10}.  The quarternion representation
can be translated into rotations in four-dimensional space.
There is a long history between the Lorentz group and the four-dimensional
rotation group.  It would be interesting to see what the quarternion
representation of polarization optics will add to this history between
those two similar but different groups.
\par

As for earlier applications of the two-by-two representation of the Lorentz
group, we note the vector representation by Fedorov~\cite{fedo70,fedo79}.
Fedorov showed that it is easier to carry out kinematical calculations
using his two-by-two representation.  For instance, the computation of
the Wigner rotation angle is possible in the two-by-two
representation~\cite{bk06jpa}.  Earlier papers on group theoretical
approaches to polarization optics include also those on Mueller
matrices~\cite{dargys12} and on relativistic kinematics and polarization
optics~\cite{pellat92}.

\section{Geometry of the Poincar\'e Sphere}\label{poincs}

We now have the four-vector $\left(S_0, S_3, S_1, S_2\right)$, which is
Lorentz-transformed like the space-time four-vector $(t, z, x, y)$ or the
energy-momentum four-vector of Eq.(\ref{mom11}).  This Stokes four-vector
has a three-component subspace $\left(S_3, S_1, S_2\right)$, which is like
the three-dimensional Euclidean subspace in the four-dimensional Minkowski
space.  In this three-dimensional subspace, we can introduce the spherical
coordinate system with
\begin{eqnarray}
&{}&  R = \sqrt{S_3^2 + S_1^2 + S_2^2} \nonumber\\[1ex]
&{}&  S_3 = R\cos\xi,       \nonumber \\[1ex]
&{}&  S_1 = R ( \sin\xi) \cos\delta, \qquad S_2 = R (\sin\xi) \sin\delta .
\end{eqnarray}
\par

\begin{figure}
\centerline{\includegraphics[scale=0.25]{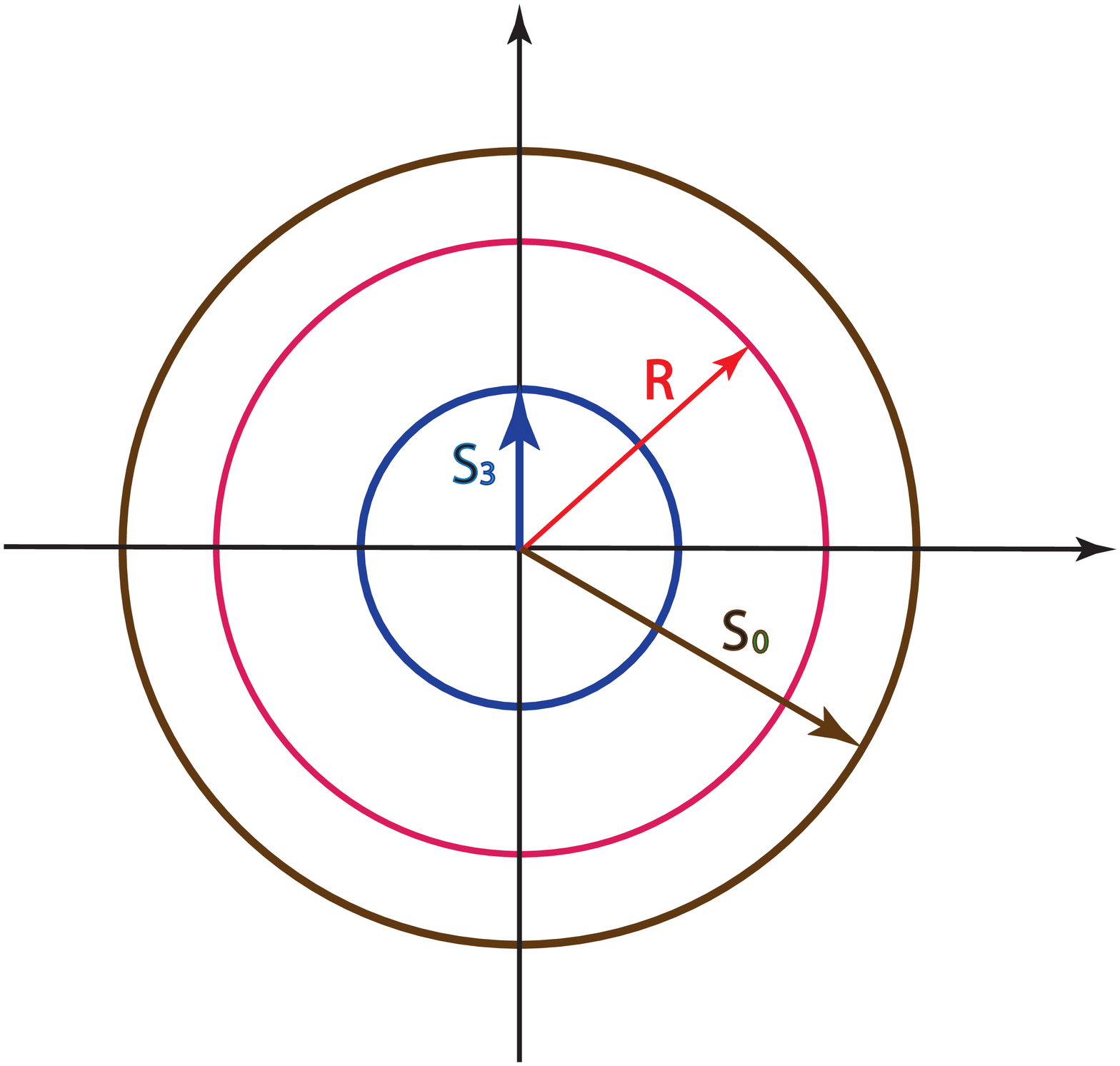}
\hspace{3mm}
\includegraphics[scale=0.25]{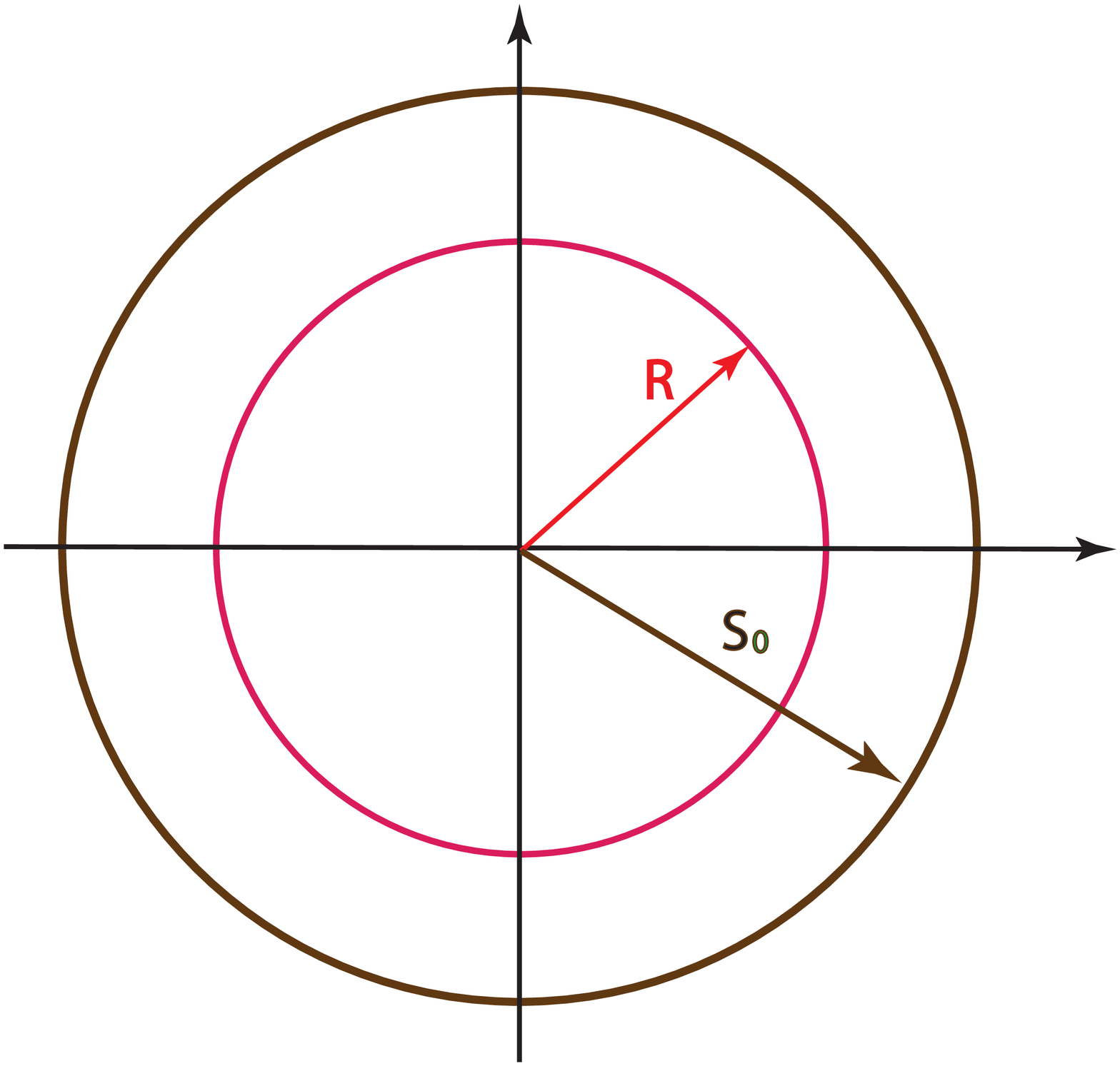}}
\caption{Radius of  the Poincar\'e sphere.  The radius $R$ takes its
maximum value $S_0$ when the decoherence angle $\chi$ is
zero.  It becomes smaller as $\chi$ increases.  It becomes minimum
when the angle reaches $90^o$.  Its minimum value is $S_3$ as is
illustrated in the left figure.  The degree of polarization is maximum
when $R = S_0$, and is minimum when $R = S_3$.  According to
Eq.(\ref{s3}), $S_3$ becomes $0$ when $a = b$, and the minimum value
of $R$ becomes zero, as is indicated in the right figure.  Its maximum
value is still $S_0$.  This maximum radius can become larger because
$b$ becomes larger to make $a = b$. }\label{poincs22}
\end{figure}

The radius $R$ is the radius of this sphere, and is
\begin{equation}\label{radius11}
 R = \frac{1}{2} \sqrt{(a^2 - b^2)^2 + 4(ab)^2\cos^{2}\chi} .
\end{equation}
with
\begin{equation}\label{s3}
S_3 = \frac{a^2 - b^2}{2}.
\end{equation}
This spherical picture is traditionally known as the Poincar\'e
sphere~\cite{azzam77,born80,bros98}.  Without loss of generality,
we assume $a$ is greater than $b$, and $S_3$ is non-negative.
In addition, we can
consider another sphere with its radius
\begin{equation}
S_0 = \frac{a^2 + b^2}{2} ,
\end{equation}
according to Eq.(\ref{stokes11}).
\par
The radius $R$ takes its maximum value $S_0$ when $\chi = 0^o$.  It
decreases and reaches its minimum value, $S_3$, when $\chi = 90^o$.
In terms of $R$, the degree of polarization given in Eq.(\ref{pdegree})
is
\begin{equation}
f = \frac{R}{S_0} .
\end{equation}
This aspect of the radius R is illustrated in Fig.~\ref{poincs22} (left).
The minimum value of $R$ is $S_3$ of Eq.(\ref{radius11}).
\par


Let us go back to the four-momentum matrix of Eq.(\ref{mom11}).  Its
determinant is $m^2$ and remains invariant.   Likewise, the determinant
of the coherency matrix of Eq.(\ref{cocy22b}) should also remain invariant.
The determinant in this case is
\begin{equation}
 S_{0}^2 - R^2 = (ab)^2 \sin^2\chi .
\end{equation}
This quantity remains invariant.  This aspect is shown on the last row
of Table~\ref{tab33}.
\par

Let us go back to Eq.(\ref{sq11}).  This matrix changes the relative
magnitude of the amplitudes $a$ and $b$.  Thus, without loss of
generality we can study the Stokes parameters with $a = b$.  The coherency
matrix then becomes
\begin{equation}\label{cocy55}
C = a^2 \pmatrix{1 & (\cos\chi)e^{-i\delta} \cr
(\cos\chi)e^{i\delta} & 1},
\end{equation}
Since the angle $\delta$ does not play any essential roles, we can let
$\delta = 0$, and write the coherency matrix as
\begin{equation}\label{cocy56}
C = a^2 \pmatrix{1 & \cos\chi \cr \cos\chi & 1}.
\end{equation}

\par
Then the minimum radius $S_3 = 0$, and $S_0$ of Eq.(\ref{stokes11})
and $R$ of Eq.(\ref{radius11}) become
\begin{equation}
S_0 = a^2, \qquad R = a^2 (\cos\chi),
\end{equation}
respectively.  The Poincar\'e sphere becomes simplified to that of
Fig.~\ref{poincs22} (right). This Poincar\'e sphere allows $R$ to decrease
to zero.

\par
The determinant of the above two-by-two matrix is
\begin{equation}
     a^4 \left(1 - \cos^2\chi\right) = a^4\sin^2\chi .
\end{equation}

Since the  Lorentz transformation leaves the determinant invariant, the
change in this $\chi$ variable is not a Lorentz transformation.  It is of
course possible to construct a larger group in which this variable plays
a role in a group transformation~\cite{bk06jpa}, but we are in this paper
more interested in its role in a particle gaining a mass.  With this
point in mind, let us diagonalize the coherency matrix of Eq.(\ref{cocy55}).
Then it takes the form
\begin{equation}\label{cocy66}
a^2 \pmatrix{1 + \cos\chi & 0 \cr 0 & 1 - \cos\chi}
\end{equation}
This form is the same as the four-momentum matrix given in Eq.(\ref{mat31}).
There we were not able to associate the variable $\chi$ with any known
physical process or symmetry operations of the Lorentz group.  Fortunately,
in this section, we noted that this variable comes from the degree of
decoherence in polarization optics.

\par

\section*{Concluding Remarks}
In this paper, we noted first that the group of Lorentz transformations
can be formulated in terms of two-by-two matrices.  This two-by-two formalism
can also be used for transformations of the coherency matrix in polarization
optics consisting of four Stokes parameters.
\par
Thus, this set of the four parameters is like a Minkowskian four-vector
under four-by-four Lorentz transformations.  In order to accommodate all
four Stokes parameters, we noted that the radius of the Poincar\'e sphere
should be allowed to vary from its maximum value to its minimum,
corresponding to the  fully and  minimal coherent cases.

\par
As in the case of the particle mass, the decoherence parameter in the
Stokes formalism is invariant under Lorentz transformations.  However,
the Poincar\'e sphere, with a variable radius, provides the mechanism
for the variations of the decoherence parameter.  It was noted that
this variation gives a physical process whose mathematics corresponds
to that of the mass variable in particle physics.

\par
As for polarization optics, the traditional approach has been to
work with two polarizer matrices like
\begin{equation}
\pmatrix{1 & 0 \cr 0 & 0}, \qquad \pmatrix{0 & 0 \cr 0 & 1} .
\end{equation}
We have replaced these two matrices by one attenuation matrix of
Eq.(\ref{atten}).  This replacement enables us to formulate the
Lorentz group for the Stokes parameters~\cite{hkn97}.  Furthermore,
this attenuation matrix makes it possible to make a continuous
transformation from one matrix to another by adjusting the
attenuation parameters in optical media.  It could be interesting
to design optical experiments along this direction.

\section*{Acknowledgments}

This paper is in part based on an invited paper presented by one of
the authors (YSK) at the Fedorov Memorial Symposium:
International Conference "Spins and Photonic Beams at Interface,"
dedicated to the 100th anniversary of F.I.Fedorov (1911-1994)
(Minsk, Belarus, 2011).  He  would like to thank Professor Sergei
Kilin for inviting him to the conference.
 \par
In addition to numerous original contributions in optics, Fedorov
wrote a book on two-by-two representations of the Lorentz group based
on his own research on this subject.  It was, therefore, quite
appropriate for him (YSK) to present a paper on applications of the
Lorentz group to optical science.  He would like thank Professors
V. A. Dluganovich and M. Glaynskii for bringing to his attention the
papers and the book written by Academician Fedorov, as well as their
own papers.

\begin{appendix}

\section*{Appendix}

In Sec.~\ref{poincw}, we listed four two-by-two matrices whose repeated
applications lead to the most general form of the two-by-two matrix $G$.
It is known  that every $G$ matrix can be translated into a four-by-four
Lorentz transformation matrix through~\cite{naimark54,knp86,hkn97}
\begin{equation}\label{trans44a}
\pmatrix{t' + z' \cr x' - iy'  \cr x' + iy' \cr t' - z'} =
\pmatrix{\alpha\alpha^{*} & \alpha \beta^{*} &
\beta\alpha^{*} & \beta \beta^{*} \cr
\alpha \gamma^{*} & \alpha \delta^{*} &
\beta \gamma^{*} & \beta \delta^{*} \cr
\gamma \alpha^{*}  & \gamma \beta^{*} &
\delta \alpha^{*} & \delta \beta^{*} \cr
\gamma \gamma^{*} & \gamma \delta^{*} &
\delta \gamma^{*} & \delta \delta^{*}}
\pmatrix{t + z \cr x - iy  \cr x + iy \cr t - z} ,
\end{equation}
\par

and
\begin{equation}\label{trans44b}
\pmatrix{t \cr z  \cr x \cr y } =  \frac{1}{2}
\pmatrix{1 & 0 & 0 & 1 \cr 1 & 0 & 0 & -1 \cr 0  & 1 & 1 & 0 \cr
0 & i & -i & 0}
\pmatrix{t + z \cr x - iy  \cr x + iy \cr t - z} .
\end{equation}

\par
These matrices appear to be complicated, but it is enough to study the
matrices of Eq.(\ref{herm11}) and Eq.(\ref{symm11}) to cover all the
matrices in this group.  Thus, we give their four-by-four equivalents
in this appendix.
\begin{equation}
Z(\delta) = \pmatrix{e^{i\delta/2} & 0 \cr 0 & e^{-i\delta/2}}
\end{equation}
leads to the four-by-four matrix
\begin{equation}
\pmatrix{1 & 0 & 0 & 0 \cr 1 & 0 & 0 & 0 \cr
      0  & 1 & \cos \delta &  -\sin\delta \cr
      0 & 0 & \sin\delta & \cos\delta}.
\end{equation}
\par
Likewise,
\begin{equation}
B(\eta) = \pmatrix{e^{\eta/2} & 0 \cr 0 & e^{-\eta/2} }
 \rightarrow
\pmatrix{\cosh\eta &  \sinh\eta & 0 & 0 \cr
    \sinh\eta & \cosh\eta & 0 &  0 \cr
   0  & 0 & 1 & 0 \cr   0 & 0 & 0 & 1},
\end{equation}

\begin{equation}
R(\theta) = \pmatrix{\cos(\theta/2) & -\sin(\theta/2) \cr
   \sin(\theta/2) & \sin(\theta/2) }
\rightarrow
\pmatrix{1 & 0 & 0 & 0 \cr
    0 & \cos\theta & -\sin\theta &  0 \cr
   0  & \sin\theta & \cos\theta & 0 \cr   0 & 0 & 0 & 1},
\end{equation}
and

\begin{equation}
S(\lambda) = \pmatrix{\cosh(\lambda/2) & \sinh(\lambda/2) \cr
  \sinh(\lambda/2) & \sinh(\lambda/2) }
\rightarrow
\pmatrix{\cosh\lambda & 0 & \sinh\lambda & 0 \cr
    0 & 1 & 0 &  0 \cr
   \sinh\lambda  & 0 & \cosh\lambda & 0 \cr
     0 & 0 & 0 & 1}.
\end{equation}
\end{appendix}

\end{document}